\documentclass[a4paper,onecolumn,twoside,fleqn,10pt,runningheads]{article}
\usepackage{amsmath}
\usepackage{authblk}
\usepackage[svgnames]{xcolor}
\definecolor{blue}{RGB}{0,0,255}
\usepackage{caption}
\usepackage[font={color=blue},figurename=Fig.,labelfont={it}]{caption}
\usepackage{bm}
\usepackage{subcaption}
\usepackage[varg]{txfonts}
\usepackage{graphicx}
\usepackage{pdfpages}
\usepackage{wasysym}
\usepackage{rotating}
\usepackage{hyperref}
\usepackage{indentfirst}
\usepackage{float}
\usepackage{lipsum} 
\usepackage{setspace} 
\usepackage{indentfirst}
\usepackage[english]{babel}
\usepackage[switch,columnwise]{lineno}
\usepackage[left=2cm,right=2cm,top=2cm,bottom=2cm]{geometry}

\hypersetup{
    colorlinks=true,                          
    linkcolor=blue, 
    citecolor=blue, 
    urlcolor=blue,
    linktoc=all
}

\newcommand{\eg}{\textit{eg }}
\newcommand{\cf}{\textit{cf }}

\usepackage{fancyhdr}
\begin{document}
\title{The Effect of Lunar Declination on CO$_{2}$ degassing from Central Italian Apennines}
\author[1]{P. Zuddas}
\author[2]{F. Lopes}

\affil[1]{Sorbonne Universit\'e, CNRS, METIS, F75005, Paris, France}			
\affil[2]{Universit\'e Paris Cité, Institut de Physique du globe de Paris, CNRS UMR 7154, F-75005 Paris, France}

\date{\today}
\maketitle
\ \\
\newpage
\ \\

\abstract {The periodical degassing from CO$_{2}$ over-pressured reservoirs may have serious consequences for the environment making urgent understanding the processes and forecasting the frequency. Prediction though needs methods that depends from temporal and spatial properties of hydro-chemical and physical reservoir characteristics that unfortunately are often lacking. We have analyzed surface emissions of CO$_{2}$ attributed to over-pressured CO$_{2}$-rich reservoirs in the Central Italian Apennines a zone characterized by significant periodical CO$_{2}$ degassing. Here aquifers are hosted in Mesozoic limestone with high pCO$_{2}$ groundwater and travertine deposits. We analyzed a 10-year temporal series and found that in the Apennines, CO$_{2}$ flux and aquifer fluid composition are correlated with the lunar tides specific to the geographic zone. In particular, our study reveals that low CO$_{2}$ flux corresponds with low lunar tidal potential values. We found a similar trend for dissolved calcium and water alkalinity, while pH values display a linear correlation with tidal cycles. The forces associated with tidal potentials are not capable of fracturing rock. However, they can, under certain conditions, drive the flow of fluids in over-pressured reservoirs, triggering sub-surface fluid movements that in turn modify the water–rock reactivity. In the central Apennines, these movements result in increased dolomite dissolution and an eventual return to calcite equilibrium. In this case, dolomite dissolution breaks the rock releasing calcium into ground water, which leads to calcite equilibrium and in turn to the formation of significant quantities of travertine and the concomitant release of CO$_{2}$ in the atmosphere.}

\section{Introduction\label{sec:01}}
	The response of aquifers to Earth and Lunar tide strains has long been observed and recognized in several situations for more than one century (\eg \cite{klonne1880,meinzer1928}) and recently used to provide specific estimation storage and permeability of the aquifers (\cf \cite{allegre2016}). The potential impact of these strains has also been proposed as responsible for changes in stress and pore pressure in porous and in water-saturated formations (\cf \cite{mcmillan2019}).  However, despite Lunar tides are ubiquitous, the potential impacts of tides have never been evaluated in over pressured reservoirs. These reservoirs may occur naturally in particular geological situations or in low carbon strategies such as the geological sequestration of CO$_{2}$ or in the more traditional CO$_{2}$-enhanced recovering when CO$_{2}$ is injected into oil field to increase energy production (\cf \cite{lake2019}).   The distribution of CO$_{2}$ emission, in the Central Italian Apennines corresponds the formation of reservoirs fed by fluids rising through processes that imply the production and accumulation of CO$_{2}$ in the Earth's upper crust (\eg \cite{uysal2009,tamburello2018}). Similarly, the geological CO$_{2}$ storage, a critical technology for environmental energy, produce oversaturated fluids conditions that may generated a CO$_{2}$ plume in the subsurface (\eg \cite{middleton2020}). The mechanisms that determine this process have complex and nonlinear interactions with geological properties (permeability, depth, thickness) and have been tentatively explained by the interaction between CO$_{2}$-saturated and CO$_{2}$-oversaturated aquifers (\eg \cite{giammanco2008,chiodini2020}). It is today difficult to predict the frequency of such variations using hydrogeological models because aquifer models assume that conditions are 'perfectly confined' or 'purely unconfined', whereas boundary conditions vary for regional-scale aquifers (\eg \cite{hsieh1987,galloway1988, roeloffs1996,christenson2011,allegre2016,wang2018}). Real temporal chemical measurements directly reflect these variations (\eg \cite{wang2018}) and furthermore take into account possible low tectonic stress and periodical lunar tide oscillations that locally reshape the aquifers (\eg \cite{jacob1940,bredehoeft1967,robinson1971,morland1984,hsieh1987,dean1994,rogie2000,merritt2004,chiodini2010,craig2017,wang2018}).
	
	In our study, we have used the chemical composition records on springs and well in the Central Apennine area (Central Italy) collected over a 10-year period (2009-2018) by Chiodini et al. (2020). We have calculated the CO$_{2}$ emissions in this area confirming earlier estimates that indicate that the quantity of CO$_{2}$ released into the atmosphere from 2009-2018 in this zone is equivalent to the total amount of CO$_{2}$ released by Mount Etna (Italy), the Earth's premier CO$_{2}$-producing volcano over the same period.  This amount is by over 1800 kt of CO$_{2}$, equivalent to a CO$_{2}$ flux of 1011 mol yr-1 that should correspond to 10\% of the overll Earth flux for this range of time (\eg \cite{allard1991,chiodini2010}). In this paper we report the effect of these external forces on the chemical composition of the aquifer sampled providing a possible mechanism able to explain the observed variation of CO$_{2}$ and of the fluid chemical composition.

\section{Data and Methods\label{sec:02}}	
\begin{figure}[H]
         \centering \includegraphics[width=0.8\columnwidth]{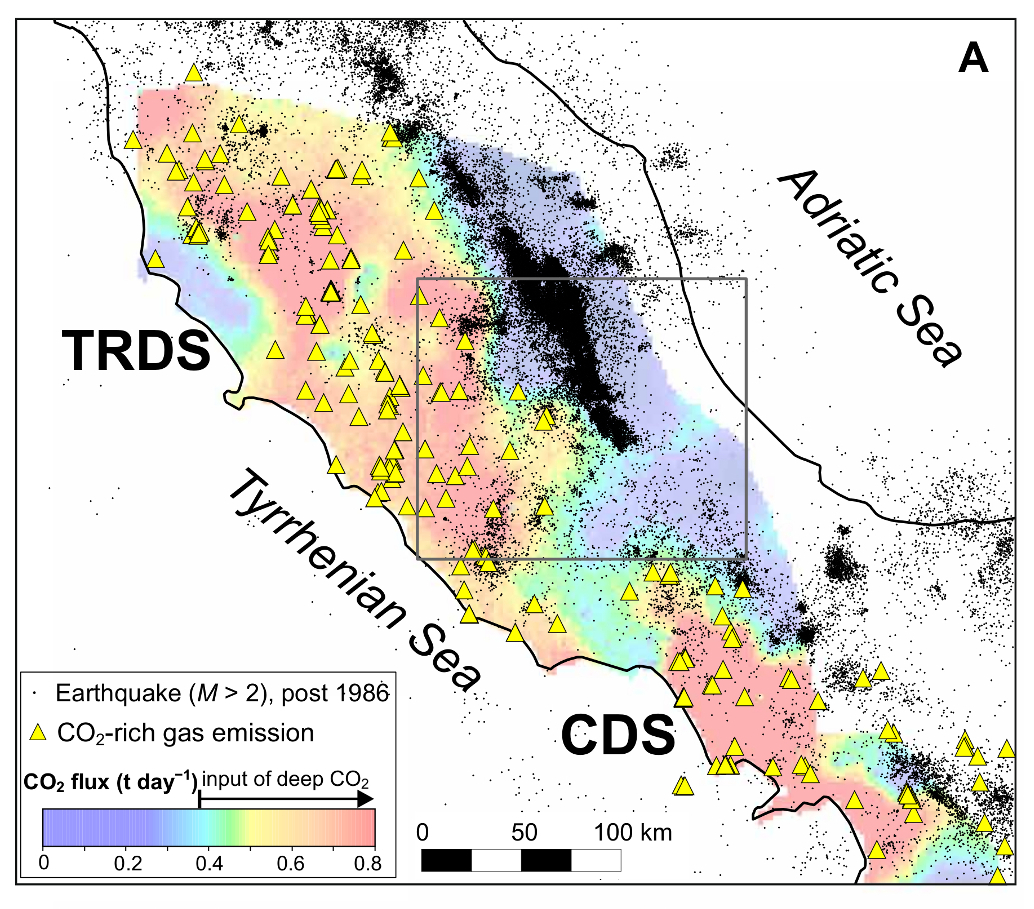}	
         \caption{Map of CO$_{2}$  degassing and location of the investigated aquifer (from Chiodini et al. \cite{chiodini2020}).}
         \label{fig:01}
\end{figure}

	The Italian Apennine region is home to several aquifers hosted in Mesozoic limestone and has frequent and significant gas emissions and numerous soda springs and wells with high pCO$_{2}$ groundwater and travertine deposits at the surface. Crustal thickness in the central Apennines is approximately 35 km, heat flow is normal and gravity anomalies are negative. The current tectonic deformation field is characterized by an extension in the axial zone of the Apennine chain and by a contraction responsible for periodic earthquakes of moderate magnitude (between 4 and 6), originating mainly in the upper crust (\eg \cite{bonfanti2012}). They are generated by active lithospheric thrusts beneath the Apennine watershed and are generally associated with low-angle normal faults accommodating uplift of basement culminations. The large amount of CO$_{2}$ released into the atmosphere is most probably controlled by diffusive processes and the regional hydrogeological features of the two main degassing areas (Fig. \ref{fig:01}). The zone under scrutiny covers 700 km$^2$ and is also characterized by numerous natural springs and wells that release CO$_{2}$ oversaturated water at the rate of hundreds of liters per second. The water releases CO$_{2}$ into the atmosphere and produces large travertine deposits that have been quarried since ancient roman times, over 2000 years ago. Field data used in this study (major and minor chemical fluid composition and dissolved CO$_{2}$) was collected from the 36 main springs and wells of the two large reservoirs, sampled between April 2009 and December 2018 (\eg \cite{chiodini2020}).

	We have recalculated pCO$_{2}$ partial pressure using field alkalinity and pH data and the overall water composition, using the PHREEQC software and thermodynamic database (\cf \cite{parkhurst2013}). Springs, gas vents and travertine are potentially controlled by regional hydrology and meteoric water. The hydrological recharge is located in the Mesozoic limestone layers that overlay a thick and relatively impermeable cover of Oligocene–Neogene clastic sediments. Spring distribution is dictated by water circulation in the Mesozoic limestone, where water chemical composition is between Ca–SO$_4$ and Ca(Mg)–HCO$_3$, with very limited amounts of HCO$_3$, Na–Cl and Na–SO$_4$.

\newpage
	
	To estimate the Lunar tides, we have considered that the Earth is spherically symmetrical, nonrotating and has self-gravity with elastic and isotropic properties in an initial equilibrium state. The Moon and the Sun are considered celestial bodies that produce a tidal effect on the Earth. Assuming $M$ = Earth masse and $m$ = Moon masse, every point $P$ at the Earth's surface is subject to a force $\gamma$ equal to:	
\begin{equation}
	\gamma = G.M.m \dfrac{\textbf{PA'}}{(PA')^{3}},
	\label{eq:01}
\end{equation}
	
	where $\dfrac{\textbf{PA'}}{(PA')^{3}}$ is the vector distance between the point $P$ at the Earth's surface and the Moon's center of gravity $A'$, while $G$ is the gravitational constant of Newton. The Earth is subject to an attraction force, $\gamma_E$, from the Moon equal to:	
\begin{equation}
	\gamma = \dfrac{G.m \textbf{AA'}}{(AA')^{3}},
	\label{eq:02}
\end{equation}
	
	where $A$ is the Earth's center of gravity, $\dfrac{\textbf{AA'}}{(AA')^{3}}$ is the vector distance between the Earth's center of gravity and the Moon's center of gravity. The difference between the Moon's attraction on $P$ and the Earth's attraction on the Moon, also called Earth tide force, $f$, is the difference between $\gamma$ and $\gamma_E$. The gravitational force $f$ can be expressed as a function of its potential $w$:
\begin{equation}
	w = G.m.[\dfrac{1}{\textbf{PA'}} - \dfrac{-\textbf{AP}*\textbf{AA'}}{(AA')^{3}}].
	\label{eq:03}
\end{equation}	

	Since the position of point $P$ on the Earth's surface is a function of the latitude $z$, we have expressed the Earth's tide potential $w$ using Legendre polynomials:
\begin{equation}
	w = \dfrac{G.m}{D}*\sum_{n=0}^{\infty} (\dfrac{R}{D})^{n}*P_{n} (\cos z),
	\label{eq:04}
\end{equation}		

	Where $D$ is the distance between the Earth's center of gravity and the Moon's center of gravity, $R$ is the distance between the given point $P$ and the Earth's center of gravity and $P_n$, is the Legendre polynomial function. The Earth's tidal potential $w$ can be approximated to the second order Legendre polynomial term:	
\begin{equation}
	w = \dfrac{G.m R^{2}*(3.\cos^{2}z - 1)}{2*D^{3}},
	\label{eq:05}
\end{equation}

	where lunar ephemeris and effective latitude positions were provided by the Institut de Mécanique Céleste et de Calcul des Éphémérides (IMCCE\footnote{http://vo.imcce.fr/webservices/miriade/?forms}). We have calculated the attraction force generated by the Moon on the Earth ($w$), also called lunar tidal potential or tidal oscillations, using an average latitude of 43$^{\circ}$ (\eg \cite{stewart2007}). 
	
\section{Results\label{sec:03}}	
\begin{figure}[H]
         \centering \includegraphics[width=1\columnwidth]{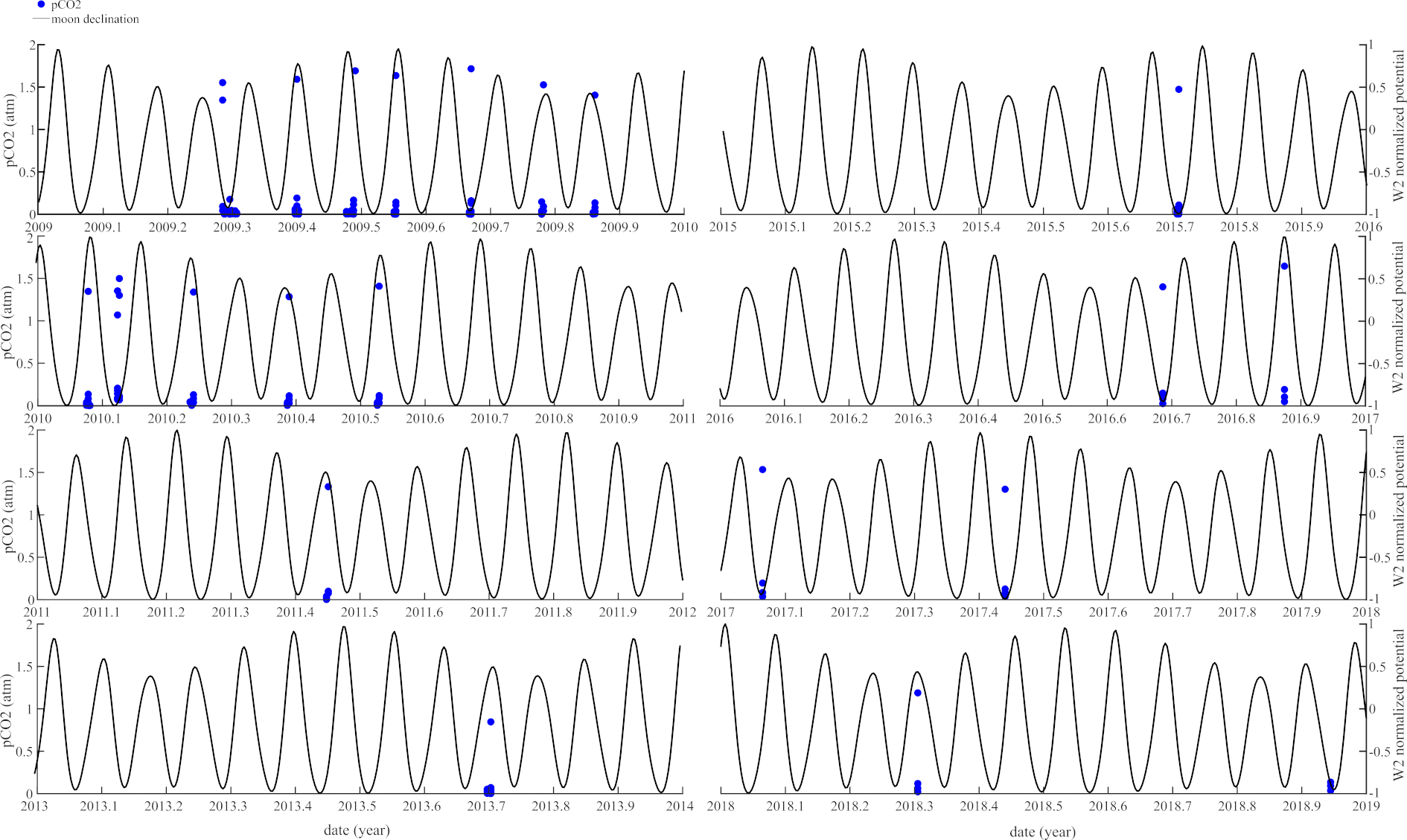}	
         \caption{ Evolution of pCO$_2$ partial pressure and Moon tide potential estimated by Eqn. \ref{eq:05} during the 10 years of investigation.}
         \label{fig:02}
\end{figure}

	The principal lunar tide oscillations estimated by our mechanical calculations show a periodicity of 27.555 days. The periodicity results from the variation in the distance between the Earth and the Moon in the central Apennines. The lunar tidal potential responsible for the attraction force on the Earth calculated by Eqn. \ref{eq:05} is reported in Fig. \ref{fig:02}. We found that the minimum tidal potential during the 10 years of investigation is linearly correlated to the minimum pCO$_2$ partial pressure values observed in the region. We were surprised by the existence of such a strong correlation, given that there is no clear mechanistic relationship between pCO$_2$ partial pressure and tidal potentials.
	
\begin{figure}[H]
	    \centering
		\includegraphics[width=0.9\columnwidth]{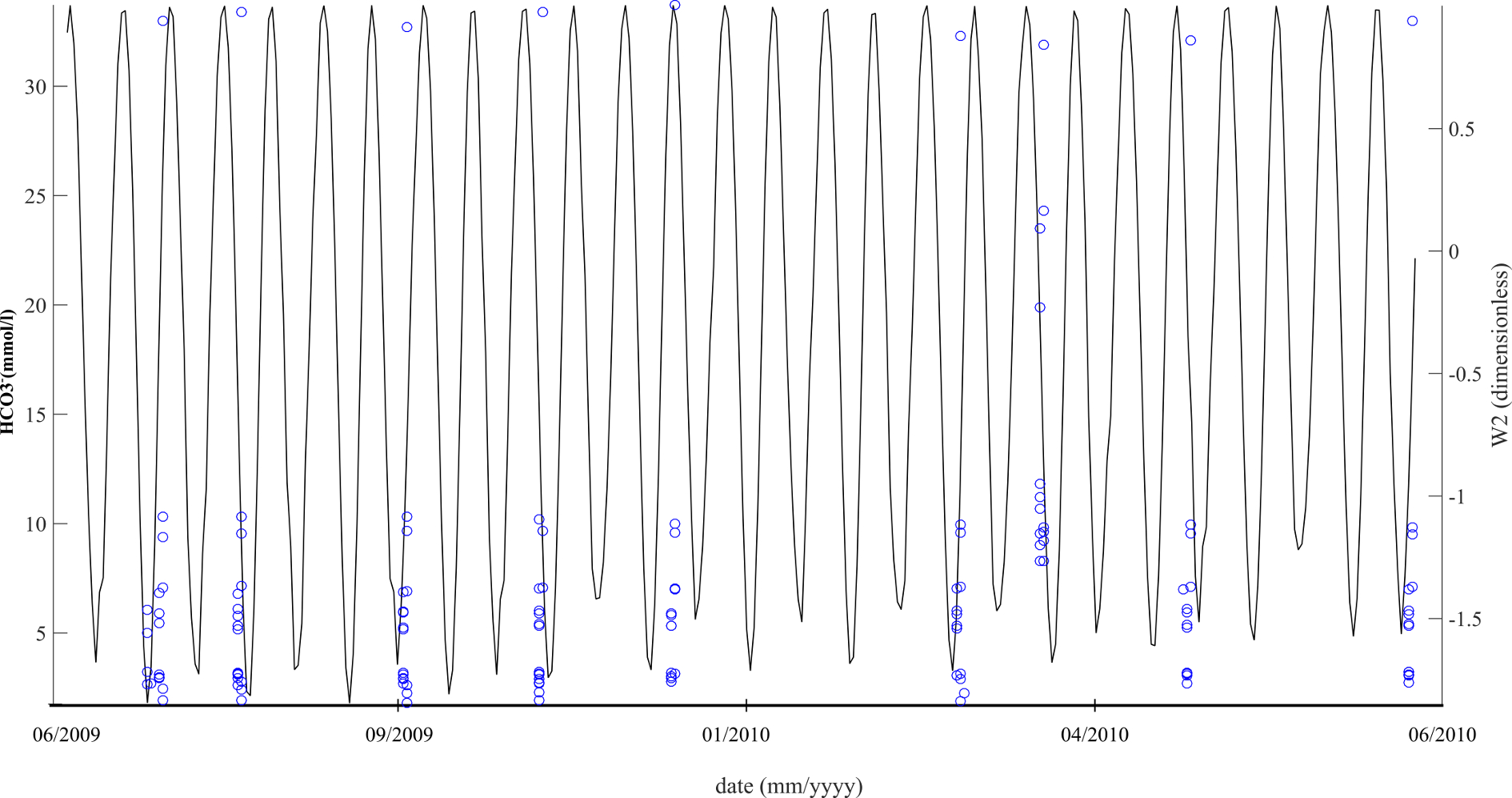}	
         \caption{Evolution of calcium concentration (red dots) as a function of time for the year: June 2009-June 2010. The continuous black line corresponds to the continuous W potentials estimated by Eqn. \ref{eq:05}.}      
		\label{fig:03}
\end{figure}

	We focused our attention on the data set gathered from June 2009 - June 2010 because it was the largest and densest of the 10-year period. Fig \ref{fig:03} illustrates that 95\% of the minimum calcium concentration corresponds to the minimum Lunar potential while only 5\% of the maximum calcium concentration value corresponds to the maximum Lunar potential. It is not possible to identify a continuous function between tidal potential variations and minimum calcium concentrations using one-year data sets because those sets do not contain any intermediate values. We can safely say however, based on our data sets, that minimum calcium concentrations are indeed related to weak Lunar tide potentials while higher calcium values are related to high Lunar potential values. 	
	
\begin{figure}[H]
	    \centering
   	    \includegraphics[width=0.9\columnwidth]{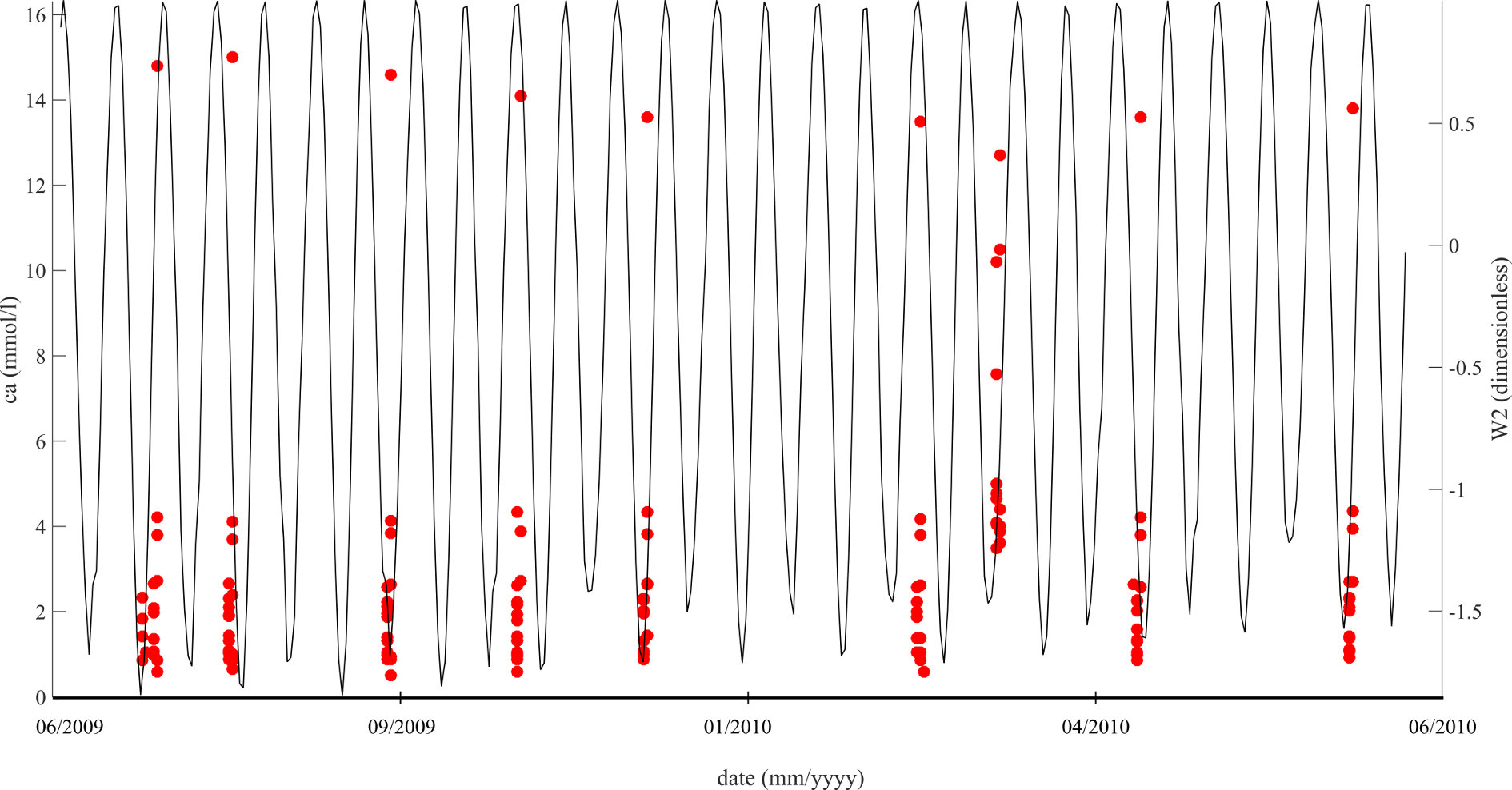}	
         \caption{Evolution of alkalinity (blue dots) as a function of time for the year: June 2009-June 2010. The continuous black line corresponds to the continuous $w$ potentials estimated by \ref{eq:05}.}
		\label{fig:04}
\end{figure}	

\begin{figure}[H]
	    \centering
		\includegraphics[width=0.9\columnwidth]{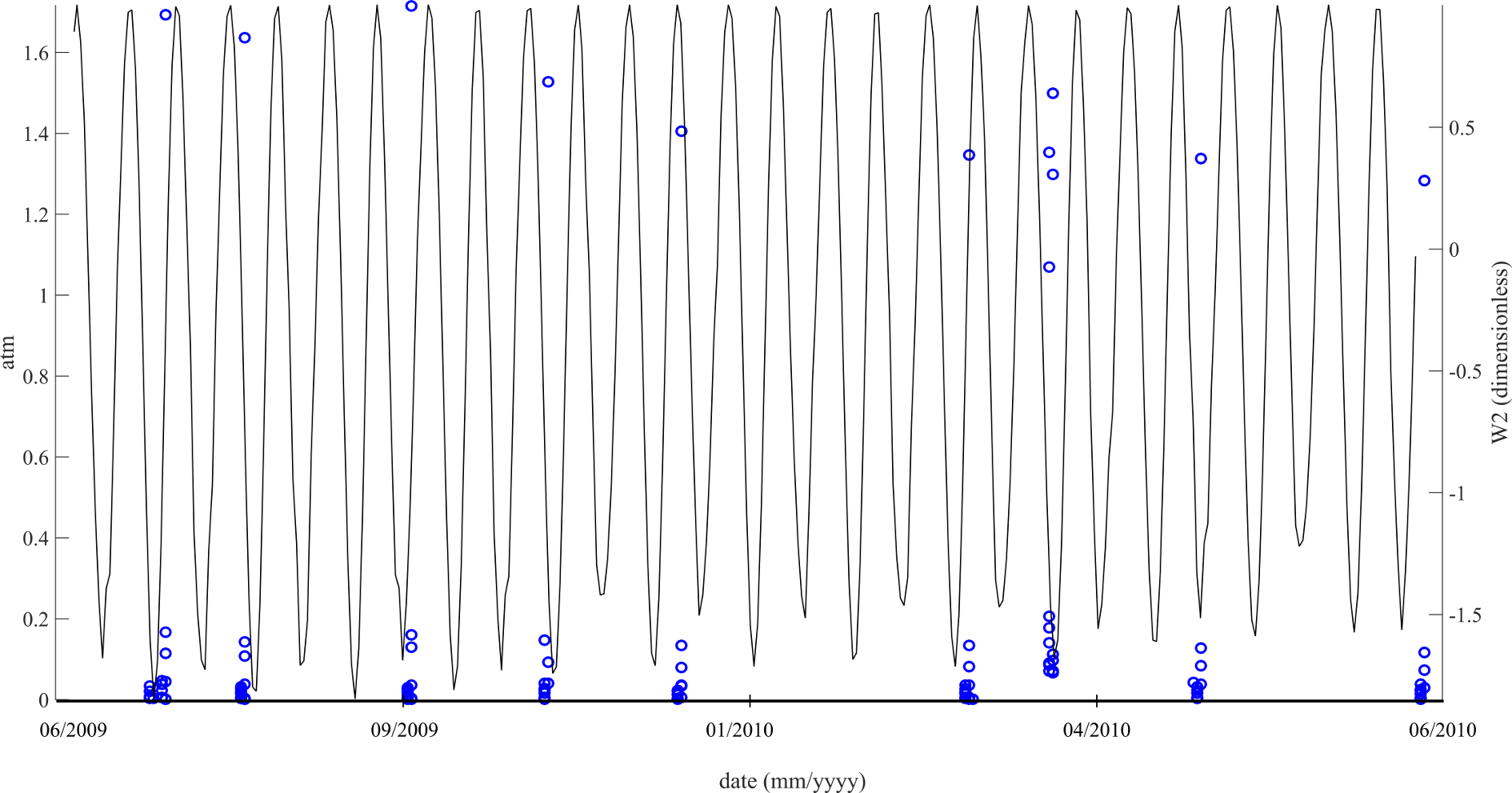}	
	    \caption{Evolution of pCO2 partial pressure (blue dots) as a function of time for the year: June 2009-June 2010. The continuous black line corresponds to the continuous $w$ potentials estimated by Eqn.  \ref{eq:05}.}
		\label{fig:05}
\end{figure}	
	
\begin{figure}[H]
	    \centering
		\includegraphics[width=0.9\columnwidth]{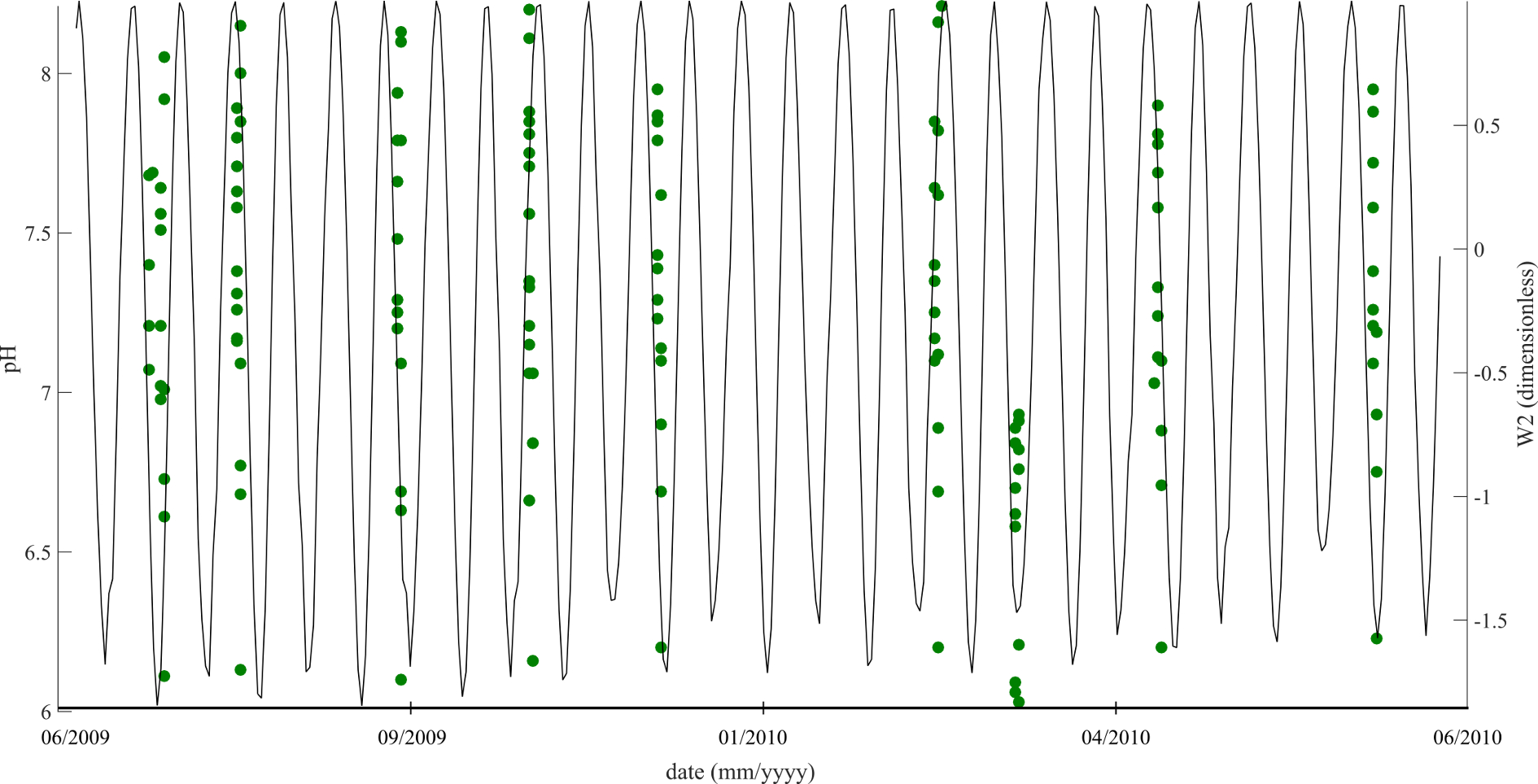}	
	    \caption{Evolution of pH (green dots) as function of time for the year: June 2009-June 2010. The continuous black line corresponds to the continuous $w$ potentials estimated by Eqn. \ref{eq:05}.}
		\label{fig:06}
\end{figure}

	Fig. \ref{fig:04} shows that variations in alkalinity as a function of time also correlate to tidal potentials. Similarly, because intermediate values are lacking, it is also not possible to identify a continuous function for alkalinity. Fig. \ref{fig:05}  shows that when plotting the pCO$_2$ partial pressure variations as a function of the one-year time-range, 95\% of the pCO$_2$ values correspond to the minimum value of estimated potentials while the remaining 5\% correspond to the minimum value. Low pCO$_2$, close to 0.1 atm, corresponds to a weak Lunar potential while higher values, close to 1.4 atm, are associated with strong Lunar potential. These observations suggest that CO$_2$ emissions in the zone may be related to Lunar tidal forces in a discontinuous and apparently non-linear relationship. We considered that is possible that pCO$_2$ variations may reflect changes in chemical reactivity in the aquifer and if that is indeed the case, it should be reflected in fluid pH variations. In fact, variations in fluid pH globally reflect the complex process of water-rock interaction because pH participates in both carbonate and silicate mineral dissolution and precipitation reactions. Fig. \ref{fig:06} illustrates that pH significantly varies between a minimum of 6.2 and a maximum of 8.2. However, unlike calcium, alkalinity and pCO$_2$ partial pressure, pH values are linearly correlated to lunar potential values during the 27.555 time-period. This indicates the existence of a water-rock interaction process responsible for CO$_2$ release. 	

	To explain the possible reason of such a linear relation, we have estimated the mineral saturation indexes (SI) of the water and found that sampled fluids are close to equilibrium with respect to calcite given that SI$_{calcite}$ is between -0.7 and +0.3 with a mean value of -0.1. It is likely that the principal process that controls groundwater composition is calcite dissolution in near-equilibrium conditions. Calcite dissolution may affect the water composition of our system and influence observed pCO$_2$ partial pressure values (\cf. \cite{chiodini2020}). The collected ground waters are under-saturated with respect to gypsum or anhydrite (CaSO$_4$. 2H$_2$O, CaSO$_4$ respectively) although it is possible that gypsum or anhydrite dissolution does take place in springs with high SO$_4$ content. When gypsum or anhydrite dissolves, it can facilitate dolomite dissolution because of the common ion effect (\cf \cite{morse1990}). This in turn could allow for increased calcite precipitation maintaining calcite in near-equilibrium conditions. Additional processes such as halite dissolution from evaporate formation may effect groundwater composition in the studied zone, given that clay-water equilibria may contribute to the observed continuous pH variations.

	However, tidal forces are too weak to fracture rock compared to tectonic forces (\eg \cite{mauk1973}). We have estimated the vertical force acting on a unit mass at the average latitude for the investigated zone and found that the average force is close to 5.6*10$^{-7}$ N.kg$^{-1}$. Given the pressure gradient associated with tidal potential, the pressure between the surface and the first 10-15 km (i.e. the transition between ductile and brittle crust) is approximately 10 Pa for an average rock density of 2.4 kg.m$^{-3}$. This value is lower by several orders of magnitude than the 0.1-0.20 MPa required to fracture carbonate rock in tensile stress (\eg \cite{schultz1996}). 10 Pa of tidal pressure may however drive fluid flow in stressed localized areas, increasing the movement of subsurface fluids (\eg \cite{mcmillan2019}). Groundwater at different depths may therefore be effected by tidal forces (\eg \cite{stillings2021}) and carry with it the gas accumulated in buried highly pressurized aquifers as it travels towards the surface. At the fault planes, fluid pressure is reduced and local aquifers may connect favoring CO$_2$ release (\eg \cite{frondini2019}) with consequent carbonate precipitation, in agreement with the following the classical reaction path:
\begin{equation}
	\textrm{Ca}^{2+} + 2\textrm{HCO}^{-}_{3} = \textrm{CaCO}^{-}_{3} + \textrm{H}_{2}\textrm{O} + \textrm{CO}_{2}\uparrow
	\label{eq:06}
\end{equation}	

		The formation of calcium carbonate deposits evidenced by the large quantities of travertine in the zone and the significant release of CO2 point to disequilibrium conditions generated by a reduction in CO2 pressure (\eg \cite{girault2018,chiodini2020}). When tidal potential is high and the Moon is close, aquifer depths are highest and fluids are oversaturated with respect to calcite; whereas when tide potential is low, fluids are close equilibrium with calcite.  Although lunar tides exert only weak stress on rock (\eg \cite{metivier2009}) without fracturing it (\eg \cite{schultz1996}), they do indeed produce weak vertical fluid movements (\eg \cite{girault2018,chiodini2020}) causing ex-solution from CO2-rich waters. This process, similar to the mechanical bubbling that occurs when you shake a bottle of carbonated soda, modifies the system's overall chemical equilibrium (\eg \cite{lix2020}) and generates fluid releasing processes. The observed variations in central Apennine CO$_2$ release may correspond to gas separation and the ascent of deep CO$_2$ through fluid movements triggered by the lunar declination. 
	
	As confirmed by our investigation, tides are harmonic functions characterized by amplitudes and phases that simultaneously act as stressors on the subsurface, the contribution of lunar tides on the groundwater response can now be disentangle. We propose that systematically exploring the influence of subsurface properties on the tidal transfer function, the relationship between the amplitude and the phase of the stressors and the resulting chemical transport reactivity responsible for CO$_2$ releasing could be potentially predicted.	
	
\section{Conclusions\label{sec:04}}	
Our observational scale work shows the relation between lunar tides and CO$_2$ releasing in over-pressured CO$_2$ aquifers. The pressurized carbonate fluids rise to the surface through pre-existing faults producing variable amounts of CO$_2$ that are potentially released into the atmosphere. We propose that carbonate mineral dissolution under calcite near-equilibrium conditions may be responsible for the observed CO$_2$ variations in the Central Apennines and that changes in fluid dynamics can be triggered by periodical tidal potential variations. 

The analysis of long-term time series of CO$_2$ discharge and chemical fluid composition variations show that periodical low values of CO$_2$ released are correlated to the minimum lunar periodical tides of the zone. The combination of tidal potentials with traditional reactive transport description constraining poroelastic and hydraulic parameter space can be potentially integrated in future predictive modelling of the CO2 releasing from natural and artificial over pressured CO2 reservoirs.
		
\paragraph{Acknowledgments}
This research was partially found by Sorbonne University and Institut de Physique du Globe de Paris joint program Emergence. We thank Drs J.L. Le Mouël and G. Chiodini for helpful discussions during the writing time of the work.   

\newpage	
\bibliographystyle{ieeetr}
\bibliography{appenini}
\end{document}